\documentclass[conference,compsoc]{IEEEtran}

\usepackage[utf8]{inputenc}
\usepackage[T1]{fontenc}

\newcommand\sectionremark[1]{%
  \vspace{-10pt}\noindent\textit{{#1}}\vspace{10pt}
}

\begin{document}

\title{ESiWACE2 Services: RSE collaborations in Weather and Climate}


%
\author{\IEEEauthorblockN{Gijs van den Oord\IEEEauthorrefmark{1},
Victor Azizi\IEEEauthorrefmark{1},
Alessio Sclocco\IEEEauthorrefmark{1},
Georges-Emmanuel Moulard\IEEEauthorrefmark{2},
David Guibert\IEEEauthorrefmark{2},\\
Jisk Attema\IEEEauthorrefmark{1},
Erwan Raffin\IEEEauthorrefmark{2},
Ben van Werkhoven\IEEEauthorrefmark{1},
\IEEEauthorblockA{\IEEEauthorrefmark{1}Netherlands eScience Center, Amsterdam, the Netherlands}
\IEEEauthorblockA{\IEEEauthorrefmark{2}Center for Excellence in Performance Programming, Atos Bull, Rennes, France}
}}

\maketitle

\begin{abstract}
We present the collaborative model of ESiWACE2 Services, where Research Software Engineers (RSEs) from the Netherlands eScience Center (NLeSC) and Atos offer their expertise to climate and earth system modeling groups across Europe. Within 6-month collaborative projects, the RSEs intend to provide guidance and advice regarding the performance, portability to new architectures, and scalability of selected applications. We present the four awarded projects as examples of this funding structure.  
\end{abstract}


\section{Introduction}

Weather and climate models are large and complex applications that experience a tension field between investments to enhance the numerical representation of the underlying physics, by e.g. resolving more processes, increasing spatio-temporal resolution or enhancing the statistics of a forecast ensemble, and on the other hand investments to adapt the software to the latest hardware architectures~\cite{gmd-11-1799-2018}. The latter investments are generally viewed within the community as a necessary burden to increase model performance and hardware utilisation, and therefore enable the aforementioned model accuracy improvements. Ideally, the code optimizations are minimally intrusive and allow domain scientists to explore the impact of new experimental features on the model results.

However, after a long period of relative stability, the computing infrastructure is rapidly evolving and diversifying, with more change on the horizon~\cite{heldens2020landscape}. The above principle of minimal changes to the source code is no longer maintainable as Fortran compilers are unable to efficiently map the data layout and compute pipeline onto these new hardware architectures. Adapting existing software to the latest computing platforms and exploiting the full capability of the architecture therefore requires specific expertise and extensive changes at the source code  level~\cite{vanwerkhoven2020lessons}. As such, continued development of weather and climate models requires new collaborations between experts from different fields~\cite{gmd-12-4425-2019}.

The ESiWACE2 Center of Excellence\footnote{http://www.esiwace.eu} aims to support the exascale preparations for the weather and climate modelling community in Europe. Within this project, we have set up a service, targeting the community at large, including marine, atmospheric, surface and ice modeling groups. The service establishes short collaboration projects between research software engineers (RSEs) and model development groups, with the goal of providing guidance, engineering, and advice to improve model efficiency and port models to existing and upcoming computing infrastructures. 

We have realized this by setting up a call for project proposals. The proposals are reviewed on technical feasibility by the RSEs, and on scientific merit and relevance by an external scientific review committee. When found eligible, the project is granted 6 person months of in-kind contribution by the RSEs from the Netherlands eScience Center and Atos. We have organized the call such that researchers can participate with minimal effort: filling in a simple form with a concise description of the requested work is sufficient. Detailed planning, descriptions of technologies or deep scientific motivation were not required in order to attract smaller, less specialized modeling groups as well. We did however require participants to be owner of the software, to invest time in guiding the RSEs during their work and provide benchmark cases, and to genuinely consider adopting the developed code in their main development branch. The first four of these projects are now proceeding and generating the first results.

%
%
In this paper, we present the case of the ESiWACE2 Service model as a way to fund RSE activities and create new collaborations between RSEs and research groups. Because the projects are relatively short, the work focuses on exploratory porting, finding and resolving performance bottlenecks, giving advice, and providing expertise rather than porting an entire code base to a new architecture. Nevertheless, the work of the RSE is expected to be a stepping stone towards exascale deployment of the software and the knowledge transfer could have a lasting impact on the development team's perspective on code optimization and hardware utilization of their application.  

\section{Case Studies}

This section presents the ongoing work in the four projects that were granted
in the in the ESiWACE2 Service 1 call, which run in the year 2020. One week in-person visits by one of the RSEs to the institute of the principal investigator of each project were originally planned as part of the project, but these visits have been cancelled due to the covid-19 pandemic.

\subsection{FESOM2}
\sectionremark{in collaboration with the Alfred Wegener Institute for Polar and Marine Research (AWI), NLeSC, and Atos}
 
The Finite-Element Sea Ice and Ocean Model FESOM 2.0~\cite{fesom} is a finite-volume ocean and sea ice model using unstructured global (or regional) meshes. The grid consists of prismatic cells and may have local refinements and a varying number of active vertical layers depending on the bathymetry. FESOM is part of the AWI earth system model and has couplings to ECHAM6 and OpenIFS via the OASIS framework. Its MPI scaling \cite{fesomoptim} and sea-ice solver method \cite{fesomice} have been subject of extensive optimization efforts. The main goal of this project is to explore the benefits of porting FESOM2 to GPUs. After a performance and code analysis and discussion with the FESOM team, we have decided to focus on the expensive tracer advection routines. We are investigating both a manual porting into tuneable CUDA-C kernels~\cite{kerneltuner}, as well as the insertion of OpenACC directives in the original Fortran code.  

\subsection{DALES}
\sectionremark{in collaboration with Centrum Wiskunde \& Informatica (CWI), NLeSC, and Atos}

The Dutch Atmospheric Large Eddy Simulation (DALES)~\cite{dales} is a high-resolution three-dimensional solver focused on boundary layer turbulence and cloud physics. It uses a rectangular grid of finite extent and periodic boundary conditions in the horizontal. The main goals of the project are to generally improve parallel scaling and memory throughput of the numerical schemes in the code. The proposed optimisations include introducing an iterative multi-grid solver for the pressure equation (instead of the current FFT-based approach), overlapping communication and computation during halo exchanges, using a more cache-friendly blocked data and loop structure, and experiment with single-precision prognostic fields. After discussing with the main developers of DALES, we have decided to focus on the pressure equation solver, the MPI optimisations, and the single-precision version of the code. We benchmark our results and timings with a simple case and a more realistic setup featuring many physical processes and a large domain extent. Initial tests show better scaling of DALES using the iterative solver when running on $\mathcal{O}(1000)$ cores.

\subsection{EMAC}
\sectionremark{in collaboration with the Cyprus Institute, Atos and NLeSC}

Embedded within the modular climate coupling framework messy2 \cite{messy2}, the ECHAM/MESSy Atmospheric Chemistry model EMAC~\cite{emac} describes chemical interactions in the atmosphere, including sources from ocean biochemistry, land processes and anthropogenic emissions. This computationally intensive code has been ported to GPUs using CUDA~\cite{alvanos2019accelerating} to achieve large speedups. The goal of our work is to reduce the memory footprint of the code in the device code, and as such allow more MPI tasks to be concurrently run on the same hardware. This will allow the model to handle an order of magnitude more complex chemistry, targeting the main organics mechanism, in atmospheric models and enable multiyear high-resolution simulations with EMAC. We currently have succeeded in significantly reducing the memory consumption by reducing stack allocations, and we are currently investigating the use of CUDA streams to further overlap compute and data transfers~\cite{vanwerkhoven2014performance}.

\subsection{OBLIMAP2}
\sectionremark{in collaboration with KNMI and Atos}

OBLIMAP-2.0~\cite{oblimap} is a fast, distributed climate–ice sheet model coupling code that includes online embeddable mapping routines. It can be used to evaluate atmospheric forcings and geothermal heat fluxes upon an ice sheet, allowing intermediate coordinate system projections between these data sources. The goal of the RSEs is to reduce the memory footprint of the code by improving its distribution over parallel tasks and secondly resolve the I/O bottleneck by implementing parallel NetCDF reading and writing. This will improve the multi-node scaling of OBLIMAP and establish it as a candidate ice coupling library for large-scale, high-resolution climate models. After a first phase in which the latest version code was moved to GitHub, the code is deployed on the 
Atos Bull Sequana XH2000 supercomputer, equipped with AMD Rome (7742), and performance profiling is ongoing.

\section{Conclusions}

With the first set of projects ongoing, we may already conclude that the collaboration with RSEs shines new light on the awarded applications. Together with the engineers, the modeling groups rethink their stance in the tension field between application performance, code maintainability and model complexity. Vice versa, the RSEs adopt knowledge about earth system modeling and the associated HPC aspects. Where some applications have already shown speedup, other projects are still exploring suitable technologies. 

We will organize new calls for project proposals in the coming years for the duration of the ESiWACE2 project and will further evaluate the funding instrument at the end of the project.
Nevertheless, it is clear that this funding strategy stimulates RSEs and HPC experts to be outward-looking and stimulate knowledge transfer towards smaller, less specialized modeling groups across the EU.

\section*{Acknowledgments}

ESiWACE2 has received funding from the European Union’s Horizon 2020 research and innovation programme under grant agreement No 823988.


\bibliographystyle{IEEEtran}
\bibliography{rse-2020-esiwace}

\end{document}